\documentclass{article}
\usepackage{graphicx}

\title{An O(1) algorithm for implementing the LFU cache eviction scheme}

\author{Prof. Ketan Shah \and Anirban Mitra \and Dhruv Matani}

\usepackage{datetime}
\newdate{date}{16}{08}{2010}
\date{\displaydate{date}}

\begin{document}
\maketitle

\begin{abstract}
Cache eviction algorithms are used widely in operating systems, databases and other systems that use caches to speed up execution by caching data that is used by the application. There are many policies such as MRU (Most Recently Used), MFU (Most Frequently Used), LRU (Least Recently Used) and LFU (Least Frequently Used) which each have their advantages and drawbacks and are hence used in specific scenarios. By far, the most widely used algorithm is LRU, both for its $O(1)$ speed of operation as well as its close resemblance to the kind of behaviour that is expected by most applications. The LFU algorithm also has behaviour desirable by many real world workloads. However, in many places, the LRU algorithm is is preferred over the LFU algorithm because of its lower run time complexity of $O(1)$ versus $O(\log n)$. We present here an LFU cache eviction algorithm that has a runtime complexity of $O(1)$ for all of its operations, which include insertion, access and deletion(eviction).

\end{abstract}




\clearpage

\section{Introduction}
The paper is organized as follows.
\begin{itemize}
\item A description of the use cases of LFU where it can prove to be superior to other cache eviction algorithms
\item The dictionary operations that should be supported by a LFU cache implementation. These are the operations which determine the strategy's runtime complexity
\item A description of how the currently best known LFU algorithm along with its runtime complexity
\item A description of the proposed LFU algorithm; every operation of which has a runtime complexity of $O(1)$
\end{itemize}

\section{Uses of LFU}

Consider a caching network proxy application for the HTTP protocol. This proxy typically sits between the internet and the user or a set of users. It ensures that all the users are able to access the internet and enables sharing of all the shareable resources for optimum network utiliization and improved responsiveness. Such a caching proxy should try to maximize the amount of data that it can cache in the limited amount of storage or memory that it has at its disposal [\cite{KS00,Z04,SSV99}].

Typically, lots of static resources such as images, CSS style sheets and javascript code can very easily be cached for a fairly long time before it is replaced by newer versions. These static resources or "assets" as programmers call them are included in pretty much every page, so it is most beneficial to cache them since pretty much every request is going to require them. Furthermore, since a network proxy is required to serve thousands of requests per second, the overhead needed to do so should be kept to a minimum.

To that effect, it should evict only those resources that are not used very frequently. Hence, the frequently used resources should be kept at the expence of the not so frequently used ones since the former have proved themselves to be useful over a period of time. Of course, there is a counter argument to that which says that resources that may have been used extensively may not be required in the future, but we have observed this not to be the case in most situations. For example, static resources of heavily used pages are always requested by every user of that page. Hence, the LFU cache replacement strategy can be employed by these caching proxies to evict the least frequently used items in its cache when there is a dearth of memory.

LRU might also be an applicable strategy here, but it would fail when the request pattern is such that all requested items don't fit into the cache and the items are requested in a round robin fashion. What will happen in case of LRU is that items will constantly enter and leave the cache, with no user request ever hitting the cache. Under the same conditions however, the LFU algorithm will perform much better, with most of the cached items resulting in a cache hit.

Pathological behaviour of the LFU algoithm is not impossible though. We are not trying to present a case for LFU here, but are instead trying to show that if LFU is an applicable strategy, then there is a better way to implement it than has been previously published.

\section{Dictionary operations supported by an LFU cache}

When we speak of a cache eviction algorithm, we need to concern ourselves primarily with 3 different operations on the cached data.
\begin{itemize}
\item Set (or insert) an item in the cache
\item Retrieve (or lookup) an item in the cache; simultaneously incrementing its usage count (for LFU)
\item Evict (or delete) the Least Frequently Used (or as the strategy of the eviction algorithm dectates) item from the cache
\end{itemize}

\section{The currently best known complexity of the LFU algorithm}

As of this writing, the best known runtimes for each of the operations mentioned above for an LFU cache eviction strategy are as follows:
\begin{itemize}
\item Insert: $O(\log n)$
\item Lookup: $O(\log n)$
\item Delete: $O(\log n)$
\end{itemize}

These complexity values folllow directly from the complexity of the binomial heap implementation and a standard collision free hash table. An LFU caching strategy can be easily and efficiently implemented using a \emph{min heap} data structure and a hash map. The min heap is created based on the usage count (of the item) and the hash table is indexed via the element's key. All operations on a collision free hash table are of order $O(1)$, so the runtime of the LFU cache is governed by the runtime of operations on the \emph{min heap} [\cite{LCKNMCK00,OOW96,ZP01,BNMK99,BM04}].

When an element is inserted into the cache, it enters with a usage count of 1 and since insertion into a \emph{min heap} costs \emph{$O(\log n)$}, inserts into the LFU cache take $O(\log n)$ time [\cite{CLRS02a}].

When an element is looked up, it is found via a hashing function which hashes the key to the actual element. Simultaneously, the usage count (the count in the max heap) is incremented by one, which results in the reorganization of the min heap and the element moves away from the root. Since the element can move down up to \(\log n\) levels at any stage, this operation too takes time $O(\log n)$.

When an element is selected for eviction and then eventually removed from the heap, it can cause significant reorganization of the heap data structure. The element with the least usage count is present at the root of the min heap. Deleting the root of the min heap involves replacing the root node with the last leaf node in the heap, and bubbling this node down to its correct position. This operation too has a runtime complexity of $O(\log n)$.

\section{The proposed LFU algorithm}

The proposed LFU algorithm has a runtime complexity of $O(1)$ for each of the dictionary operations (insertion, lookup and deletion) that can be performed on an LFU cache. This is achieved by maintaining 2 linked lists; one on the access frequency and one for all elements that have the same access frequency.

A hash table is used to access elements by key (not shown in the diagram below for clarity). A doubly linked list is used to link together nodes which represent a set of nodes that have the same access frequency (shown as rectangular blocks in the diagram below). We refer to this doubly linked list as the \emph{frequency list}. This set of nodes that have the same access frequency is actually a doubly linked list of such nodes (shown as circular nodes in the diagram below). We refer to this doubly linked list (which is local to a particular frequency) as a \emph{node list}. Each node in the \emph{node list} has a pointer to its parent node in the \emph{freqency list} (not shown in the diagram for clarity). Hence, nodes $x$ and $y$ will have a pointer back to node $1$, nodes $z$ and $a$ will have a pointer back to node $2$ and so on...

\begin{figure}[h] 
\centering 
\includegraphics[width=10cm]{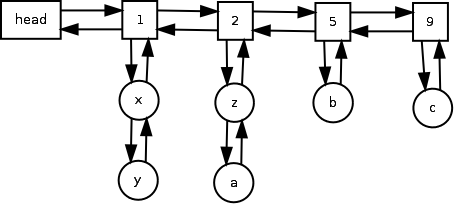}
\caption{The LFU dict with 6 elements}
\label{fig:1}
\end{figure} 

\begin{figure}[h] 
\centering 
\includegraphics[width=12cm]{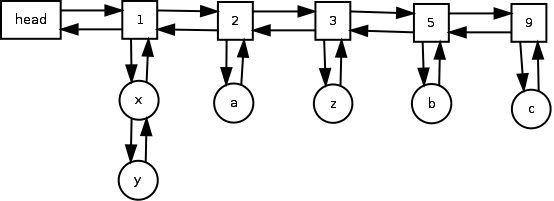}
\caption{After element with key 'z' has been accessed once more}
\label{fig:2}
\end{figure}

The pseudocode below shows how to initialize an LFU Cache. The hash table used to locate elements by key is indicated by the variable $bykey$. We use a SET in lieu of a linked list for storing elements with the same access frequency for simplicitly of implementation. The variable $items$ is a standard SET data structure which holds the keys of such elements that have the same access frequency. Its insertion, lookup and deletion runtime complexity is $O(1)$.\\


\noindent \emph{Creates a new frequency node with a access frequency value of 0 (zero)}\\
\noindent NEW-FREQ-NODE()\\
01 \hspace*{0.1in} Object o\\
02 \hspace*{0.1in} o.value $\leftarrow$ 0\\
03 \hspace*{0.1in} o.items $\leftarrow$ SET()\\
04 \hspace*{0.1in} o.prev $\leftarrow$ o\\
05 \hspace*{0.1in} o.next $\leftarrow$ o\\
06 \hspace*{0.1in} \textbf{return} o\\

\noindent \emph{Creates a new LFU Item which is stored in the lookup table \textbf{bykey}}\\
\noindent NEW-LFU-ITEM(data, parent)\\
01 \hspace*{0.1in} Object o\\
02 \hspace*{0.1in} o.data $\leftarrow$ data\\
03 \hspace*{0.1in} o.parent $\leftarrow$ parent\\
04 \hspace*{0.1in} \textbf{return} o\\

\noindent \emph{Creates a new LFU Cache}\\
\noindent NEW-LFU-CACHE()\\
01 \hspace*{0.1in} Object o\\
02 \hspace*{0.1in} o.bykey $\leftarrow$ HASH-MAP()\\
03 \hspace*{0.1in} o.freq\_head $\leftarrow$ NEW-FREQ-NODE\\
04 \hspace*{0.1in} \textbf{return} o\\

\noindent \emph{The LFU cache object is accessible via the \textbf{lfu\_cache} variable}\\
\noindent lfu\_cache $\leftarrow$ NEW-LFU-CACHE()\\

\noindent We also define some helper functions that aid linked list manipulation\\\\
\noindent \emph{Create a new new and set its previous and next pointers to prev and next}\\
GET-NEW-NODE(value, prev, next)\\
01 \hspace*{0.1in} nn $\leftarrow$ NEW-FREQ-NODE()\\
02 \hspace*{0.1in} nn.value $\leftarrow$ value\\
03 \hspace*{0.1in} nn.prev $\leftarrow$ prev\\
04 \hspace*{0.1in} nn.next $\leftarrow$ next\\
05 \hspace*{0.1in} prev.next $\leftarrow$ nn\\
06 \hspace*{0.1in} next.prev $\leftarrow$ nn\\
07 \hspace*{0.1in} \textbf{return} nn\\

\noindent \emph{Remove (unlink) a node from the linked list}\\
DELETE-NODE(node)\\
01 \hspace*{0.1in} node.prev.next $\leftarrow$ node.next\\
02 \hspace*{0.1in} node.next.prev $\leftarrow$ prev\\

Initially, the LFU cache starts off as an empty hash map and an empty frequency list. When the first element is \textbf{added}, a single element in the hash map is created which points to this new element (by its key) and a new frequency node with a value of \emph{1} is added to the frequency list. It should be clear that the number of elements in the hash map will be the same as the number of items in the LFU cache. A new node is added to 1's frequency list. This node actually points back to the frequency node whose member it is. For example, if $x$ was the node added, then the node $x$ will point back to the \emph{frequency node} $1$. Hence the runtime complexity of element insertion is $O(1)$.\\

\noindent \emph{Access (fetch) an element from the LFU cache, simultaneously incrementing its usage count}\\
\noindent ACCESS(key)\\
01 \hspace*{0.1in} tmp $\leftarrow$ lfu\_cache.bykey[key]\\
02 \hspace*{0.1in} \textbf{if} tmp \textbf{equals} \textbf{null} \textbf{then}\\
03 \hspace*{0.3in} \textbf{throw} Exception("No such key")\\
04 \hspace*{0.1in} freq $\leftarrow$ tmp.parent\\
05 \hspace*{0.1in} next\_freq $\leftarrow$ freq.next\\
06\\
07 \hspace*{0.1in} \textbf{if} next\_freq \textbf{equals} lfu\_cache.freq\_head \textbf{or}\\
08 \hspace*{0.1in} next\_freq.value \textbf{does not equal} freq.value + 1 \textbf{then}\\
08 \hspace*{0.3in} next\_freq $\leftarrow$ GET-NEW-NODE(freq.value + 1, freq, next\_freq)\\
09 \hspace*{0.1in} next\_freq.items.add(key)\\
10 \hspace*{0.1in} tmp.parent $\leftarrow$ next\_freq\\
11 \\ 
12 \hspace*{0.1in} freq.items.remove(key)\\
13 \hspace*{0.1in} \textbf{if} freq.items.length \textbf{equals} 0 \textbf{then}\\
14 \hspace*{0.3in} DELETE-NODE(freq)\\
15 \hspace*{0.1in} \textbf{return} tmp.data\\

When this element is \textbf{accessed} once again, the element's frequency node is looked up and its next sibbling's value is queried. If its sibbling does not exist or its next sibbling's value is not 1 more than its value, then a new frequency node with a value of 1 more than the value of this frequency node is created and inserted into the correct place. The node is removed from its current set and inserted into the new frequency list's set. The node's frequency pointer is updated to point to its new frequency node. For example, if the node $z$ is accessed once more (\ref{fig:1}) then it is removed from the \emph{frequency list} having the value of $2$ and added to the \emph{frequency list} having the value of 3 (\ref{fig:2}). Hence the runtime complexity of element access is $O(1)$.\\

\noindent \emph{Insert a new element into the LFU cache}\\
INSERT(key, value)\\
01 \hspace*{0.1in} \textbf{if} key \textbf{in} lfu\_cache.bykey \textbf{then}\\
02 \hspace*{0.3in} \textbf{throw} Exception("Key already exists")\\
03\\
04 \hspace*{0.1in} freq $\leftarrow$ lfu\_cache.freq\_head.next\\
05 \hspace*{0.1in} \textbf{if} freq.value \textbf{does not equal} 1 \textbf{then}\\
06 \hspace*{0.3in} freq $\leftarrow$ GET-NEW-NODE(1, lfu\_cache.freq\_head, freq)\\
07\\
08 \hspace*{0.1in} freq.items.add(key)\\
09 \hspace*{0.1in} lfu\_cache.bykey[key] $\leftarrow$ NEW-LFU-ITEM(value, freq)\\

When an element with the least access frequency needs to be \textbf{deleted}, any element from the 1\(^{st}\) \emph{(leftmost)} frequency list is chosen and removed. If this frequency list's node list becomes empty due to this deletion, then the frequency node is also deleted. The element's reference from the hash map is also deleted. Hence the runtime complexity of deleting the least frequently used element is $O(1)$.\\

\noindent \emph{Fetches an item with the least usage count (the least frequently used item) in the cache}\\
\noindent GET-LFU-ITEM()\\
01 \hspace*{0.1in} \textbf{if} lfu\_cache.bykey.length \textbf{equals} 0 \textbf{then}\\
02 \hspace*{0.3in} \textbf{throw} Exception("The set is empty")\\
03 \hspace*{0.1in} \textbf{return} lfu\_cache.bykey[ lfu\_cache.freq\_head.next.items[0] ]\\

Thus, we see that the runtime complexity of each of the dictionary operations on an LFU cache is $O(1)$.


\clearpage

\bibliographystyle{amsplain}
\bibliography{lfu}

\end{document}